\documentclass[aps,pre,twocolumn,showpacs,superscriptaddress,groupedaddress]{revtex4-1}
\usepackage{amsmath}
\usepackage{amssymb}
\usepackage{amsfonts}
\usepackage{graphicx}
\usepackage{dcolumn}
\usepackage{bm}
\usepackage{sidecap}
\usepackage{subfloat}
\usepackage{parskip}
\usepackage{grffile}

\usepackage{color}
\usepackage{hyperref}
\usepackage{hhline}
\usepackage{mathtools}
\usepackage{graphics}
\usepackage{multirow}
\usepackage{verbatim}
\usepackage{longtable}
\usepackage{rotating}
\usepackage{setspace}
\usepackage{epsfig}
\usepackage{subfigure}
\usepackage{epstopdf}
\usepackage{gensymb}
\usepackage[normalem]{ulem}
\usepackage[table]{xcolor}

\makeatletter
\def\@eqnnum{{\normalsize \normalcolor (\theequation)}}
 \makeatother
 
\graphicspath{{./}{images/}}

\begin{document}
\title{Localization of multilayer networks by the optimized single-layer rewiring}
\author{Sarika Jalan$^{1,2}$}
\email{sarikajalan9@gmail.com} 
\author{Priodyuti Pradhan$^{1}$}
\affiliation{1. Complex Systems Lab, Discipline of Physics, Indian Institute of Technology Indore, Khandwa Road, Simrol, Indore-453552, India}
\affiliation{2. Centre for Biosciences and Biomedical Engineering, Indian Institute of Technology Indore, Khandwa Road, Simrol, Indore-453552, India}

\date{\today}

\begin{abstract}
We study localization properties of principal eigenvector (PEV) of multilayer networks. Starting with a multilayer network corresponding to a delocalized PEV, we rewire the network edges using an optimization technique such that the PEV of the rewired multilayer network becomes more localized. The framework allows us to scrutinize structural and spectral properties of the networks at various localization points during the rewiring process. 
We show that rewiring only one-layer is enough to attain a MN having a highly localized PEV. Our investigation reveals that a single edge rewiring of the optimized MN can lead to the complete delocalization of a highly localized PEV. This sensitivity in the localization behavior of PEV is accompanied by a pair of almost degenerate eigenvalues. This observation opens an avenue to gain a deeper insight into the origin of PEV localization of networks. Furthermore, analysis of multilayer networks constructed using real-world social and biological data show that the localization properties of these real-world multilayer networks are in good agreement with the simulation results for the model multilayer network. The study is relevant to applications that require understanding propagation of perturbation in multilayer networks.
\end{abstract}


 \maketitle
\section{Introduction}
The traditional monolayer network framework offers only a limited representation of complex systems having different layers of interactions. Recent years have witnessed emergence of the multilayer network (MN) framework, which provides more accurate insights into the behaviors of complex systems possessing multiple types of relations among the same units \cite{physics_multilayer_network,multilayer_boccaletti_2014,multilayer_csarkar_2016, multilayer_pramod_2015}. For example, the collective behavior of a society, that is modeled by individuals interacting through the Facebook and Twitter social networks, can be better understood by considering a MN consisting of layers representing the network of people in each social media. The interactions within a layer (intra-layer connection) for this particular network model of a social system encode friendship relationships 
between the pairs of two people within each social media. Whereas the interactions between the layers (inter-layer connection) represent the impact of interactions in one layer on the other; for example, two people actively interacting by Facebook may lead to an increase in 
their Twitter activities as well \cite{physics_multilayer_network}. Another example of a real-world system which inherently has multiple types of relations is the brain. In the brain MN, one layer corresponds to a physical network, and another to a functional relationship among neurons \cite{neuroscience_2017}. Furthermore, the physical layer can also itself a MN in the synaptic level. Neurons can be connected by chemical or electric synapses forming a brain MN \cite{baptista_2010,chimera_2017}. Recently, Internet routing protocol IPv4 and IPv6 autonomous systems have also been analyzed through MN framework \cite{real_mux_networks_2016}. 

Furthermore, interactions among the constituents of a system provide a backbone for the sustenance of the dynamical behavior or functionality of the entire system. For instance, in the Facebook-Twitter MN, information propagates through the links in the individual layer and spread of information propagation depends on the architecture of the underlying network. Neurons in the brain interact to perform specific functions. Reconfiguration or rewiring of functional brain networks is required during the learning phases \cite{dynamic_reconfig_2011, wiring_2006}. Therefore scrutiny of network architecture is thus important as `structure affects function' and vice-versa \cite{rev_Strogatz_2001}. 

\begin{figure}[t]
\centering
\includegraphics[width=1\columnwidth]{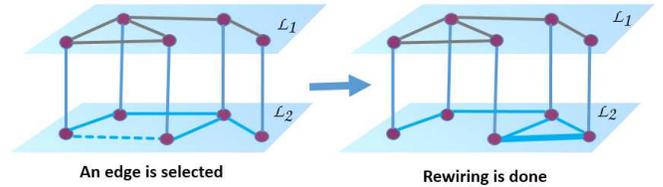}
\caption{(Color online) Schematic representation of the evolution of a multilayer network using the single-layer optimized edge rewiring scheme. A rewiring is accepted if it increases the IPR value of PEV of the multilayer networks.}
\label{fig1}
\end{figure}
The last 20 years have witnessed the development of methods and techniques to characterize various structural properties and functional activities of networks representing complex systems. Particularly, it has been reported that the eigenvector corresponding to the largest eigenvalue, the so-called principal eigenvector (PEV) of the network's adjacency matrices, provides information about both the structural and dynamical properties of the underlying systems 
\cite{Goltsev_prl2012, mieghem_epl_2012, pevec_nat_phys_2013, castellano_localization_2017, evec_localization_2017, neurons_localization_2014}. For various dynamical processes on networks, for instance, disease-spreading, the steady-state vector has been shown to be approximated using PEV of the underlying adjacency matrix \cite{Goltsev_prl2012, mieghem_epl_2012}.  To understand how individual entity is infected or how information spreads in a network in the steady state, it is sometimes enough to analyze the PEV of the corresponding adjacency matrices. The behavior of the disease spreading in the SIS model has been investigated with the help of PEV localization revealing its connection with various structural properties of the underlying monolayer networks \cite{Goltsev_prl2012, evec_localization_2017, satorras_localization2016, loc_bipartite_2017}. The localization of an eigenvector refers to a state where few components of the vector take very high values, and the rest of the components take very small values. We quantify the localization of an eigenvector using the inverse participation ratio (IPR) \cite{Goltsev_prl2012} (see also Eq. (\ref{eq_IPR})). Moretti et al. used PEV localization of the corresponding adjacency matrix to analyze the brain network dynamics \cite{brain_networks2013}. Recently, Arruda et al. extended the PEV localization concepts for MNs \cite{multilayer_disease_loc_2017} and identified that the PEV localization behaviors for MNs could be different from the monolayer networks. Specifically, in the monolayer networks, localization can happen on few nodes \cite{Goltsev_prl2012} whereas in MNs a layer can be localized \cite{multilayer_disease_loc_2017}. These investigations shed light on the properties of the networks and their relations with eigenvectors, particularly PEV. However, it remains unclear what specific structural properties the MNs should have so that they make the corresponding PEV localized. Additionally, how network structure of an individual layer affects or regulates the PEV localization of the entire MN? What role other layers of a MN play in restricting the impact on the regulating layers. Specifically, the question which we address here using the optimization technique is that what structural properties an individual layer should possess so that they correspond to a highly localized PEV of the entire MN.

In this article, we examine various structural and spectral properties of the MNs as layers evolve from a state having a completely delocalized PEV to a state having a very highly localized PEV. Our investigations reveal that the highly localized PEV of the MN for a given network size 
possesses specific structural properties, such as the presence of a hub node, high clustering coefficient, and low degree-degree correlation.
For a two layers MN, the optimization process can be implemented considering two different edge rewiring protocols; (1) by rewiring edges in both-layers or (2) by rewiring edges in only one (accessible) layer. For both the rewiring protocols, though there is an emergence of various specific structural features, the different rewiring protocols lead to a noticeable and essential difference in the spectral properties of the optimized MN structure. For the both-layers rewiring protocol, the PEV is sensitive to a single edge rewiring in the optimized MN structure as also observed in the monolayer networks \cite{evec_localization_2017}, however, interestingly, we get rid of this sensitivity of the PEV for the single-layer rewiring protocol. 

We can summarize our study as follows; Starting with an initial random MN,  where the individual layer is represented by a random monolayer network, we rewire the intra-layer edges with an optimization-based method by considering IPR value of the PEV as the fitness function. The initial random MN corresponding to the delocalized PEV evolves to an optimized structure corresponding to a highly localized state. We examine various structural and spectral properties of this optimized MN structure. Additionally, the rewiring scheme adopted here allows us to scrutinize various structural and spectral properties of the MNs at various steps of the evolution process.

We present our results for two layers, three layers, and four layers MNs. Additionally, we consider various real-world MNs constructed using empirical data taken from social and biological systems and analyze their PEV localization behaviors. 

\section{Methods} 
First, we represent a MN, $\mathcal{M}=(\mathcal{G},\mathcal{C})$ \cite{multilayer_boccaletti_2014}, where $\mathcal{G}=\{\mathcal{L}_{\alpha};\; \alpha \in \{1,2,\ldots, l\}\}$  is the family of connected monolayer network $\mathcal{L}_{\alpha} =\{V_{\alpha},E_{\alpha}\}$, where $V_{\alpha}=\{v_1^{\alpha}, v_2^{\alpha},\ldots,v_n^{\alpha}\}$ is the set of vertices and $E_{\alpha}=\{e_1^{\alpha}, e_2^{\alpha},\ldots,e_{r(\alpha)}^{\alpha}:e_{r(\alpha)}=(v_i^{\alpha}, v_j^{\alpha})\} \subseteq U_{\alpha}$ is the set of edges in the $\alpha$ layer of the MN. We define the universal set $U_{\alpha} = V_{\alpha} \times V_{\alpha}=\{(v_i^{\alpha}, v_j^{\alpha}):v_i^{\alpha}, v_j^{\alpha} \in V_{\alpha}\; \text{and}\; i \neq j\}$ which contains all possible ordered pairs of vertices excluding the self-loops and the complementary set can be defined as $E_{\alpha}^c = U_{\alpha} - E_{\alpha}=\{(v_i^{\alpha}, v_j^{\alpha}):(v_i^{\alpha}, v_j^{\alpha}) \in U_{\alpha}\; \text{and}\; (v_i^{\alpha}, v_j^{\alpha}) \notin E_{\alpha}\}$ i.e., $E_{\alpha}\cap E_{\alpha}^c=\varnothing $ and $E_{\alpha}\cup E_{\alpha}^c= U_{\alpha}$. In addition, $\mathcal{C}=\{E_{\alpha\beta}\subseteq V_{\alpha} \times V_{\beta}:\; \alpha, \beta \in \{1,2,\ldots, l \}, \alpha \neq \beta \}$ is the set of edges between $\mathcal{L}_\alpha$ and $\mathcal{L}_\beta$ layers. We refer $E_{\alpha}$ as the set of all intra-layer edges and  $E_{\alpha\beta}=\{e_1^{\alpha\beta}, e_2^{\alpha\beta},\ldots,e_n^{\alpha\beta}\}$ as the set of all inter-layer edges of $\mathcal{M}$. Here, we consider each node in one layer connected to its mirror node in the other layers of the MN, and all the layers consist of exactly the same number of nodes.

Second, we denote the adjacency matrices corresponding to $\mathcal{L}_{\alpha}$  as $A^{\alpha} \in \Re^{n \times n}$ which can be defined as $(a^{\alpha})_{ij} = 1$, if  $v_i^{\alpha} \sim v_j^{\alpha}$ and 0 otherwise. We represent degree of a node $v_{i}^{\alpha}$ as $d_{v_{i}^{\alpha}}=\sum_{j=1}^{n_{\alpha}} (a^{\alpha})_{ij}$ and the average degree of $\alpha$ layer as $\langle k_{\alpha}\rangle =\frac{1}{n}_{\alpha} \sum_{i=1}^{n_{\alpha}} d_{v_{i}^{\alpha}}$. The average degree of the MN is denoted as $\langle k \rangle = 1+\frac{\sum_{\alpha=1}^{l} \langle k_{\alpha}\rangle}{l}$. For all the model MNs, each layer has the same average degree and same number of nodes. Here, we consider two layers MN with  $\mathcal{L}_{1}=\{V_1,E_1\}$ and $\mathcal{L}_{2}=\{V_2,E_2\}$, where $|V_1|=|V_2|=n$, $|E_1|=m_1$, $|E_2|=m_2$, $|E_1^c|=\frac{n(n-1)}{2}-m_1$, $|E_2^c|=\frac{n(n-1)}{2}-m_2$ and $|E_{12}|=n$. Hence, the total number of nodes in $\mathcal{M}$ is $|V|=2n=N$ and edges $|E|=m_1+m_2+n=M$. 
The supra-adjacency matrix \cite{multilayer_boccaletti_2014} of the MN is a block matrix and can be defined as:
$$
{\bf A} = \left[
\begin{array}{cc}
 {\bf A^{1}} & {\bf \mathcal{I}}  \\ 
 {\bf \mathcal{I}} &  {\bf A^{2}}
 \end{array}\right]
$$
where $\mathcal{I}$ is an $n \times n$ identity matrix. As ${\bf A^{1}}$, ${\bf A^{2}}$ and ${\bf A}$ are real symmetric matrices, each has real eigenvalues. In addition, the networks are connected. Hence, we know from the Perron-Frobenius theorem \cite{miegham_book2011} that all the entries in the PEV of ${\bf A}$ are positive. We calculate the IPR of the MN \cite{Goltsev_prl2012} as follows:
\begin{equation} \label{eq_IPR}
Y_{x_{k}^\mathcal{M}}=\sum_{i=1}^N (x_{k})_{i}^4 
\end{equation}
where $(x_{k})_{i}$ is the $ith$ component of the orthonormal eigenvector, ${\bf x_k}$, with $1 \leq k \leq N$, of the MN.  
A delocalized eigenvector with component $[1/\sqrt{N},1/\sqrt{N},\ldots,1/\sqrt{N}]$ has the IPR value 
$1/N$, whereas the most localized eigenvector with components $[1,0,\ldots,0]$ yields an IPR value equal to $1$. For a connected MN, IPR value of the PEV lies between $ 1/N \leq Y_{x_{k}^\mathcal{M}}<1$. 

In addition, to assess the contribution of an individual layer to the IPR value of PEV of the MN, we define,
\begin{equation} \label{layer_contrbution} 
Y_{x_{1}^\mathcal{M}} =  C_{x_{1}^{\mathcal{L}_1}}  + C_{x_{1}^{\mathcal{L}_2}} 
\end{equation}
\begin{equation}\tiny
\nonumber
 Y_{x_{1}^{\mathcal{M}}} = \underbrace{(x^1_{1})_{1}^4+(x^1_{1})_{2}^4+\cdots+(x^1_{1})_{n}^4}_{\mathcal{L}_1} 
             +\underbrace{(x^2_{1})_{n+1}^4+(x^2_{1})_{n+2}^4+\cdots+(x^2_{1})_{2n}^4}_{\mathcal{L}_2} 
\end{equation}
where $(x^1_{1})_i$ and $(x^2_{1})_j$ are the $ith$ and $jth$ entry in the PEV of MN from the $\mathcal{L}_1$ and $\mathcal{L}_2$ layers respectively. 
Note that contribution from the individual layers in the IPR value of the PEV of the entire MN (represented by $C_{x_{1}^{\mathcal{L}_1}}$ \&  $C_{x_{1}^{\mathcal{L}_2}}$) and the IPR value of the PEV of layers as monolayer networks (represented by $Y_{x_{1}^{\mathcal{L}_1}}$ \& $Y_{x_{1}^{\mathcal{L}_2}}$) are different.

Starting from a connected two layers MN with each layer constructed from an Erd\"os-R\'enyi (ER) random network, we rewire the edges uniformly and independently at random with an optimization-based method. Only those edge rewirings are approved which lead to an increase in the IPR value (Fig. \ref{fig1}). We are interested in assessing various properties of the MNs during the network evolution and of those networks which have highly localized PEV, i.e., the optimized MNs. 

We first examine the impact of the optimized rewiring for the two layers MN, with both-layers and single-layer rewiring protocols, and then apply the rewiring scheme to the MNs consisting of three and four layers. For the single-layer rewiring protocol, we choose an edge $e_i^{1} \in E_1$ uniformly and independently at random from $\mathcal{L}_1$ and remove it (Fig. \ref{fig1}). At the same time, we introduce an edge in the $\mathcal{L}_1$ layer from $E_1^c$, which preserves the total number of edges during the network evolution in $\mathcal{L}_1$ and also in $\mathcal{M}$. Similarly, for the both-layers rewiring protocol, we choose a layer independently and uniformly at random from $\mathcal{M}$ and follow the same approach as adopted for the single-layer rewiring protocol for the selected layer. Note that for both the rewiring protocols, we do not rewire any edges in $E_{12}$. We remark that during the network evolution there is a possibility that an edge rewiring disconnects the corresponding layer, i.e., leads to the layer having isolated nodes which are connected only through inter-layer connections without having any intra-layer connection. To avoid this situation, we only approve those rewirings which yield the nodes in a layer connected. Further, the evolution takes place in a manner that keeps the network size fixed. 

The optimization problem can be defined as: Given an input MN $\mathcal{M}$ with $N$ vertices, $M$ edges and a function $\zeta:\Re^{N} \rightarrow \Re$, we want to compute the maximum possible value of an objective function $\zeta(\mathcal{M})$ over all the simple, connected, and undirected MN  $\mathcal{M}$. Thus, we maximize the objective function, $\zeta(\mathcal{M})$ = $Y_{x_{1}^{\mathcal{M}}}$ subject to the constraints that $\sum_{i=1}^N (x_1)_i^2 = 1$ and $0 < (x_1)_i < 1$. The first constraint simply says that the PEV of $A$ is normalized in Euclidean norm. The second constraint implicitly stipulates that the network must be connected in the optimization method. In our numerical simulation, we keep the layers, as well as the MN, connected using Depth-first search method \cite{algorithms_cormen_2009}. We perform the optimization process by applying simulated annealing method \cite{simuated_annealing_1983}. We refer the initial network as $\mathcal{M}_{init}$ and the optimized network as $\mathcal{M}_{opt}$.

\section{Results and Discussion} 
We begin the investigation by analyzing the impact of changes in the architecture of the individual layers on the PEV localization of the entire MNs.
For the layers in an MN, we consider various combinations of ER random network, Barab\'asi-Albert scale-free (SF) network, star network (STAR), and regular lattice (1D) network \cite{network_sec_2016}. The ER random network is generated with an edge probability $\langle k_{\alpha} \rangle/n_{\alpha}$, where $\langle k_{\alpha} \rangle$ is the average degree of the $\mathcal{L}_{\alpha}$ layer. The SF network is constructed using Barab\'asi-Albert preferential attachment model \cite{network_sec_2016}. 
\subsection{Localization of model multilayer network}
\begin{figure}[t]
\centering
\includegraphics[width=0.6\columnwidth]{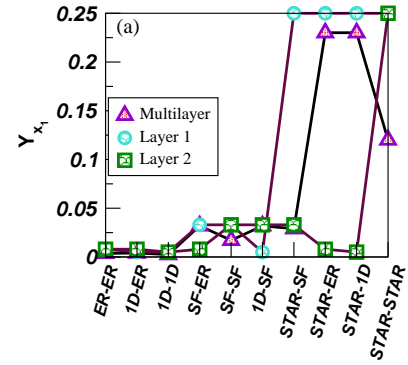}
\caption{(Color online) IPR value of monolayer and multilayer model networks for different combinations.  
The size of the monolayer networks; $n_1=n_2=200$ and $\langle k_1 \rangle =\langle k_2 \rangle =10$. Thus, size and average degree of the MN is $N=400$ and $\langle k \rangle=11$ respectively.}
\label{fig2}
\end{figure}

After multilayering of two monolayer networks, we conjecture that the IPR value of the entire MN is smaller than the maximum IPR value of the individual layers for the same number of nodes.
\begin{equation}\label{conjecture}
Y_{x_{1}^\mathcal{M}} < \max_{1 \le \alpha \le l} \{Y_{x_{1}^{\mathcal{L}_{\alpha}}}\}  
\end{equation} 
For the few combinations, multilayering may yield a high $Y_{x_{1}^\mathcal{M}}$ value and for the few combinations, the multilayering can lead to a low $Y_{x_{1}^\mathcal{M}}$ value (Fig. \ref{fig2}) however, the Eq. (\ref{conjecture}) always holds. For example, the STAR-ER and STAR-1D MNs have higher IPR values as compared to other multilayer configurations investigated here (Fig. \ref{fig2}). For the regular monolayer network (Theorem 6 \cite{miegham_book2011}), we have 
\begin{equation}\nonumber
{\bf x_1}^{\mathcal{L_{\alpha}}}=(\frac{1}{\sqrt{n}},\frac{1}{\sqrt{n}},\ldots,\frac{1}{\sqrt{n}}) 
\end{equation} 
Therefore, from Eq.~(\ref{eq_IPR}) we get $Y_{x_{1}^\mathcal{L_{\alpha}}} =\frac{1}{n}$ which corresponds to the most delocalized PEV for a network size $n$. Next, for a star monolayer network consisting of $n$ nodes with the hub node being labeled as $1$, we get the PEV as 
\begin{equation}\nonumber
{\bf x_1}^\mathcal{L_{\alpha}}=\biggl(\frac{1}{\sqrt{2}},\frac{1}{\sqrt{2(n-1)}},\ldots, \frac{1}{\sqrt{2(n-1)}}\biggr) 
\end{equation}
which yields, 
\begin{equation}\nonumber
Y_{ x_{1}^\mathcal{L_{\alpha}}}=\frac{1}{4}+ \frac{1}{4(n-1)} 
\end{equation}
For $n \rightarrow \infty$, $Y_{x_{1}^\mathcal{L_{\alpha}}} \rightarrow \frac{1}{4} \approx 0.25$. Upon multilayering two 1D monolayer networks of size $n$ and node degree $\langle k_{1} \rangle=\langle k_{2} \rangle=r$, the degree of each node of MN gets increased by one yielding the same degree to each node of the MN as ($r + 1$). Thus, 1D-1D MN network becomes regular network of $\langle k \rangle=r+1$ and $2n$ number of nodes. Therefore, PEV of the 1D-1D MN will be 
\begin{equation} \label{lower_bound}
{\bf x_1}^\mathcal{M}=(\frac{1}{\sqrt{2n}},\frac{1}{\sqrt{2n}},\ldots,\frac{1}{\sqrt{2n}})\;\;  \text{and} \;\; Y_{x_{1}^\mathcal{M}}=\frac{1}{2n}
\end{equation} 
resulting in the same contribution of each layer which is calculated from Eq.~(\ref{layer_contrbution}) as $C_{x_{1}^{\mathcal{L}_1}}=C_{x_{1}^{\mathcal{L}_2}}=\frac{1}{4n}$, and $Y_{x_{1}^{\mathcal{L}_1}}=Y_{x_{1}^{\mathcal{L}_2}}=\frac{1}{n}$ from Eq. (\ref{eq_IPR}), respectively. Therefore, both the layers contribute equally to the IPR value of the MN and the IPR value of the overall MN decreases by a factor of $1/2$ for 1D-1D MN configurations. The ER random network has a delocalized PEV for large $n$ \cite{deloc_pev}, therefore, again multilayering of two ER random networks brings upon the same contribution from both the layers in $Y_{x_1^{\mathcal{M}}}$. 

Next, if we consider STAR-1D or STAR-ER MN, the contribution $C_{x_{1}^{\mathcal{L}_2}}$ becomes very small as compared to 
$C_{x_{1}^{\mathcal{L}_1}}$. In these cases, $99.99\%$ of the contribution comes from the layer which has the STAR network for $n \rightarrow \infty$. For STAR-ER case, the PEV entry corresponding to the hub node of the STAR network has a significantly high value. On the contrary, ER random network has a delocalized PEV. After multilayering, PEV of the STAR-ER MN contains one very large entry which in combination with other tiny entries lead to a significantly high IPR value. However, for the case of STAR-STAR MN, the presence of two hub nodes leads to a decrease in the IPR value of $\mathcal{M}$ (Fig. \ref{fig2}). Similarly, for SF-ER and SF-SF networks, the presence of several hub nodes reduces the IPR value of $\mathcal{M}$. 
Following Eqs.~(\ref{conjecture}) and (\ref{lower_bound}) we get a bound for the IPR value of MNs having the same number of nodes in all the layer,
\begin{equation}\nonumber
\frac{1}{2n} \leq Y_{x_{1}^\mathcal{M}}< \max \{Y_{x_{1}^{\mathcal{L}_1}},Y_{x_{1}^{\mathcal{L}_2}}\} 
\end{equation}
In general, for $l$ layers MNs, we get 
\begin{equation}\nonumber
\frac{1}{nl} \leq Y_{x_{1}^\mathcal{M}}< \max_{1 \le \alpha \le l} \{Y_{x_{1}^{\mathcal{L}_\alpha}}\} 
\end{equation}
It is not surprising that multilayering of a delocalized monolayer network with a localized monolayer network structure leads to a higher IPR value of the MN as compared to multilayering with a delocalized monolayer network ($Y_{x_{1}^{ER-STAR}} > Y_{x_{1}^{ER-ER}}$). Additionally, it is also possible that multilayering of a localized monolayer network with another localized monolayer network (e.g., STAR-STAR) yields an IPR value which is lower than that of the localized \& delocalized (e.g., STAR-ER) multilayer network combinations (Fig. \ref{fig2}). These experiments demonstrate that PEV localization of a multilayer network can be regulated by changing topological properties of one or both of its layers.

\subsection{Layer rewiring based on simulated annealing}  
From the above experiments, we already have obtained an idea of the structural properties of an individual layers corresponding to a localized PEV state as well as how by choosing an appropriate multilayering one can make the PEV of the entire MN more localized. These investigations have been carried out for few specific network structures representing each layer of the MNs. In the following, we aim to address the issue of the PEV localization for MN having a general network architecture representing each layer. Particularly, we investigate that starting with an initial random MN, how an optimized rewiring of one or more than one layer, can build a MN having a highly localized PEV. Additionally, we investigate various structural and spectral properties of the rewired layers and those of the entire MN during the optimized evolution process at various rewiring stages.

\begin{figure}[t]
\centering
\includegraphics[width=1\columnwidth]{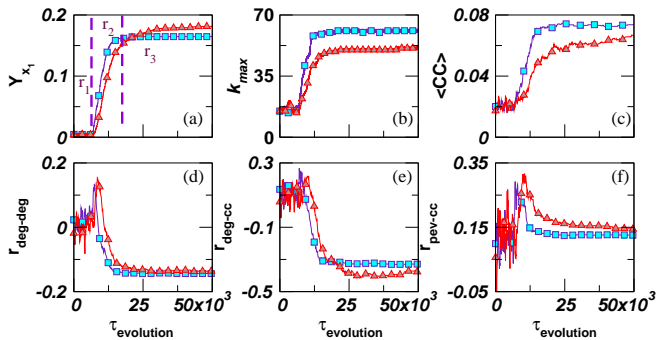}
\caption{(Color online) Optimized evolution of an initial ER-ER multilayer network for $50,000$ edge rewirings for single-layer ($\blacksquare$) and both-layers ($\blacktriangle$). Size of the MN is $N=400$, $\langle k \rangle = 7$, and $\langle k_1 \rangle =\langle k_2 \rangle$. }
\label{fig3}
\end{figure}

It can be noticed that from an initial ER-ER random MN, the optimized rewiring for both-layers, as well as for the single-layer significantly increase the IPR value (Fig. \ref{fig3}(a)) of $\mathcal{M}$. The choice of an ER-ER MN at the beginning of the evolution provides a delocalized PEV to start with \cite{deloc_pev}. During the network evolution, there are several changes in the structural and spectral properties of the network architecture of the rewired layer. For both the optimization protocols as evolution progress, IPR value of the PEV shows an increase and finally becomes saturated. Based on the nature of the increment in the IPR value, we can divide the evolution into three different regions, the slow ($r_1$), the fast ($r_2$), and the saturation ($r_3$) regions respectively (Fig. \ref{fig3}(a)). 

\begin{figure}[t]
\centering
\includegraphics[width=0.7\columnwidth]{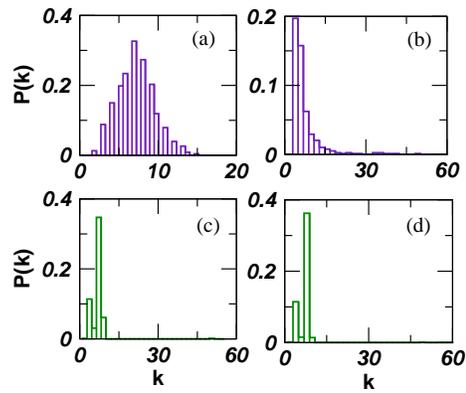}
\caption{(Color online) Degree distribution of the initial multilayer network of (a) ER-ER ($N=400$, $\langle k_1\rangle =\langle k_2 \rangle=6$) and (b) SF-SF ($N=400$, $\langle k_1\rangle =\langle k_2 \rangle=6$). (c) and (d) depict the degree distribution of the optimized MN achieved through the
both-layers rewiring protocol.}
\label{deg_dist}
\end{figure}

\begin{figure}[t]
\centering
\includegraphics[width=1\columnwidth]{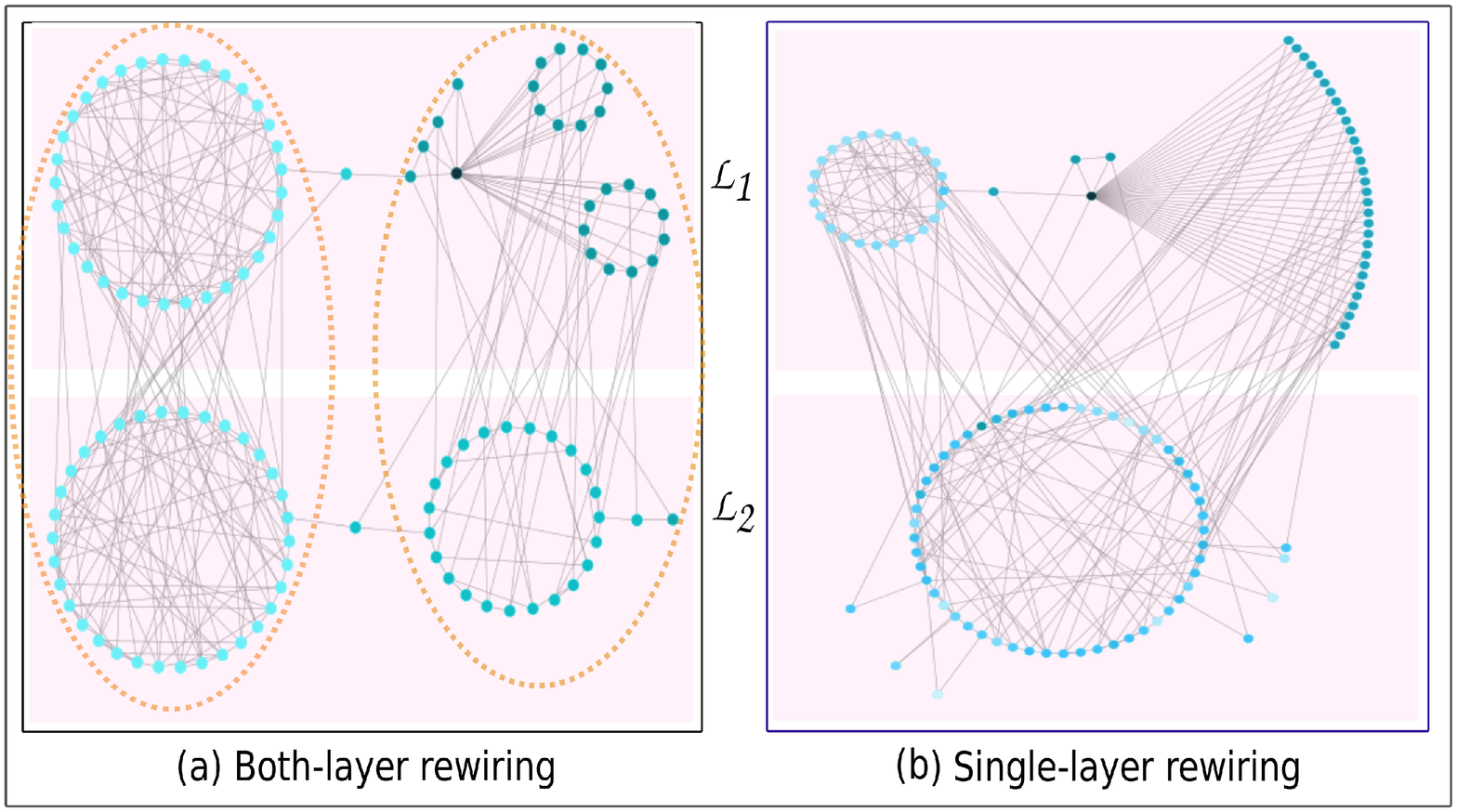}
\caption{(Color online) Cytoscape diagram of optimized MN obtained for (a) both-layers and (b) single-layer rewirings. 
For both the protocols, $N=120$ and $\langle k \rangle=3$, where $\langle k_1 \rangle =\langle k_2 \rangle$. A smaller size MN is considered here for a clear illustration of the optimized network structure.}
\label{single-both}
\end{figure}

As evolution progresses, there is a formation of the hub node (Fig. \ref{fig3}(b)) and IPR value of the PEV shows an increase which finally becomes saturated in $\mathcal{M}_{opt}$. This evolution process leads to a drastic change in the degree distribution of the final MN (Fig. \ref{deg_dist}). There exists one node in the MN coming from the smaller part of the network (Fig. \ref{single-both}) which has a very high degree as it is connected with all the nodes in that part of the network. Rest of the nodes in this part of the network has very small degrees. The other part of the layer, which does not consist the hub node, has all the nodes having degree again very small but different than those lying in the smaller part. This leads to two distinguishable peaks in the degree distribution of the optimized MN (Fig. \ref{deg_dist}). Note that instead of random initial MN, if we start with a MN having both the layers having SF topology, the final optimized network will be same as achieved for the initial MN having a random structure. We have plotted degree distribution of the final optimized MNs as well as those of the initial networks (Fig. \ref{deg_dist}). It is interesting to note that, despite the scalefree (SF) networks being more localized than the corresponding ER random networks, if we evolve a SF-SF MN using the optimization technique, the degree sequence of the final optimized structure will be the same to that achieved for optimizing the ER-ER MNs.

Additionally, $\mathcal{M}_{opt}$ has a higher average clustering coefficient ($\langle CC \rangle$) value (Fig. \ref{fig3}(c)) and low degree-degree correlation ($r_{deg-deg}$) value (Fig. \ref{fig3}(d)) as compared to those of $\mathcal{M}_{init}$ \cite{network_sec_2016}. It indicates that localization of the  PEV leads to the triangle formation in the MN structure. Similarly, the existence of a lower degree-degree correlation suggests that hub nodes are connected with lower degree nodes in individual layer leading to the MN with highly localized PEV being disassortative. 

Furthermore, to check the relation between the degree and local clustering coefficient of each node as PEV gets localized, we calculate the Pearson product-moment correlation coefficient \cite{network_sec_2016} measure of degree vector and local clustering coefficient vector ($r_{deg-cc}$) during the optimization process. It unveils that as evolution progresses the $r_{deg-cc}$ value decreases (Fig. \ref{fig3}(e)) which indicates that as the PEV gets localized, the participation of lower degree nodes is more to the cluster formation than the higher degree nodes. We measure the Pearson product-moment correlation coefficient between pairs of various other structural properties to accomplish a better understanding of the network structures. It is surprising to see that $r_{pev-cc}$ value increases (Fig. \ref{fig3}(f)) as compared to $\mathcal{M}_{init}$ as PEV gets localized. From the $r_{deg-cc}$ and $r_{pev-cc}$ values, we can also infer the correlation between degree vector and PEV ($r_{deg-pev}$) which decreases as PEV gets more localized. From these correlation measures, it is evident that lower degree nodes contribute more to the triangle formation and also contributing more to the PEV entry of the $\mathcal{M}_{opt}$. These correlation measures provide insight for possible architectures of the $\mathcal{M}_{opt}$ structure corresponding to highly localized PEV. Note that $\langle CC \rangle$ and all the correlation measures are evaluated for the entire MN.

\begin{figure}[t]
\centering
\includegraphics[width=0.8\columnwidth]{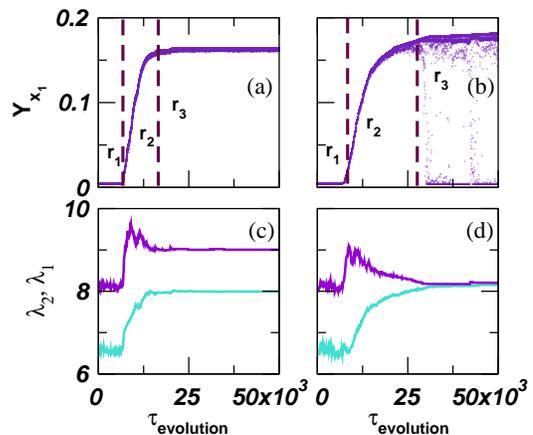}
\caption{(Color online) Changes in the IPR values ($Y_{x_1}$) with the evolution. (a) Single-layer rewiring protocol (PI) does not show IPR drops in the saturation region. (b) Both-layers rewiring protocol (PII) shows drops in the IPR value in the saturation region. (c) The behavior of largest two eigenvalues of the single-layer and (d) both-layer rewiring protocols. Here, $N=400$, $\langle k \rangle = 7$, and $\langle k_1 \rangle =\langle k_2 \rangle$.}
\label{fig6}
\end{figure}

Network visualization software reveals that the optimized layer consists of two components which are connected with each other via a single node (Fig. \ref{single-both}). One of the components in this structure contains a hub node. For both-layers rewiring protocol, we get a network structure in which one layer is similar to that obtained for the monolayer network rewiring \cite{evec_localization_2017}. However, another layer has a network structure consisting of two components of different sizes devoiding of the hub node (Fig. \ref{single-both}(a)). Various structural properties of $\mathcal{M}_{opt}$ obtained through the single-layer rewiring protocol (Fig. \ref{single-both}(b)) is qualitatively the same as that observed for the rewiring of the monolayer networks. However, for the both-layers and single-layer rewiring protocols, there is a striking (Fig. \ref{fig6}(a-b)) difference in the spectral properties in the saturation region, $r_3$. In this region, there exist several edges, rewiring which do not lead to an increase in the IPR value. If we consider rewiring of all the edges during each step of the evolution, we can notice a substantial difference between the both vs. single layer rewiring protocols of the MNs. In the $r_3$ region (Fig.~\ref{fig6}), the IPR value gets almost saturated, and there may exist only a subtle increment in the IPR value with a further evolution of the network. Although the MN in this region has the maximum IPR value, in the $\mathcal{M}_{opt}$ achieved through the both-layers rewiring protocol there exist only a few edges, for which rewiring leads to a sudden drop in the IPR value. It leads to a complete delocalization of the PEV from a highly localized state (Fig.~\ref{fig6}(b)). Thus, for both-layers rewiring, the PEV in the $r_3$ region is sensitive to a single edge rewiring as also observed for the monolayer network rewiring \cite{evec_localization_2017}.
However, for the $\mathcal{M}_{opt}$ in the $r_3$ region achieved through the single-layer rewiring protocol, there are no such sudden drops (Fig. \ref{fig6}(a)) and PEV is robust to a single edge rewiring.  
For the single-layer rewiring, the component consisting of the hub node (in the rewired layer) has a major contribution to the IPR value of the PEV of the MN, followed by the contribution from the fixed layer and the second component of the rewired layer connected to the hub-component (Fig. \ref{single-both}(b)). Similarly, for the case of both-layers rewiring, the hub-component contributes the most, followed by the contribution from the other parts of $\mathcal{M}_{opt}$ (Fig. \ref{single-both}(a)). The component containing the hub node has an overwhelming contribution in the corresponding PEV entries accompanied by an equally negligible contribution from the rest of the nodes lead to a high IPR value in the optimized structure. 

\begin{figure}[t]
\centering
\includegraphics[width=1\columnwidth]{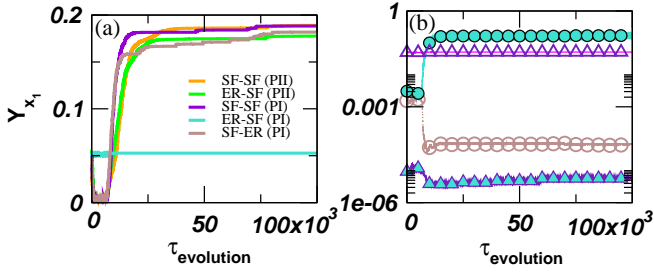}
\caption{(Color online) (a) Change in the IPR value during the evolution of MNs with various combinations of the initial networks for single-layer rewiring protocol (PI) and both-layer rewiring protocol (PII). (b) Value of $C_{x^{\mathcal{L}_1}_1}$ and $C_{x^{\mathcal{L}_2}_1}$ as 
two layers are multilayered and evolved. The initial multilayer network takes two different combinations;  ER-ER and ER-SF with $N=400, \langle k \rangle = 7$, and $\langle k_1 \rangle =\langle k_2 \rangle$. The rewiring has been done in the first layer (i.e., ER layer). The contribution
of the rewired ER layer is represented by $C_{x^{\mathcal{L}_1}_1}$. ($\bullet$) and ($\blacktriangle$) represents this contribution for ER-ER, and ER-SF multiplex networks, respectively. Similarly, the fixed layer contribution ($C_{x^{\mathcal{L}_2}_1}$) for  ER-ER and ER-SF multiplex network is depicted by ($\circ$) and ($\vartriangle$), respectively.}
\label{layer_contribution}
\end{figure}

For various combinations of the layer forming the MN (Fig. \ref{layer_contribution})(a)), for the single-layer rewiring protocol, the fixed layer restricts the IPR value of the entire MN. In Fig. \ref{layer_contribution}(b), we depict the values of $C_{x^{\mathcal{L}_1}_1}$ and $C_{x^{\mathcal{L}_2}_1}$ of $Y_{x^{\mathcal{M}}_1}$ during the network evolution for the initial MN having ER-ER and ER-SF configurations. For the ER-ER MNs, the layer which undergoes rewiring associates more weight to the PEV entries on the expense of that of the contributions from the fixed layer (Fig. \ref{layer_contribution}(b)). Both of these factors lead to a high value of $Y_{x^{\mathcal{M}}_1}$. For ER-SF MNs considered as initial networks, we observe that rewiring of the ER random layer through the optimized evolution is not sufficient to change the IPR value of PEV which is
reflected by almost a constant value of $C_{x^{\mathcal{L}_1}_1}$ and $C_{x^{\mathcal{L}_2}_1}$ 
(Fig. \ref{layer_contribution}(b)). This constant value of IPR is a consequence of the existence of the hub nodes in the fixed SF layer which imposes a restriction on the increase in $Y_{x_1^{\mathcal{M}}}$. However, for the combination of SF-ER MN, rewiring of the SF layer leads to an enhancement in the $Y_{x_1^{\mathcal{M}}}$ value (Fig. \ref{layer_contribution})(a)).

We can see that during the evolution (Fig. \ref{layer_contribution}(b)), though one layer is fixed and rewiring is performed on the other layer, changes happen in both $C_{x^{\mathcal{L}_1}_1}$ and $C_{x^{\mathcal{L}_2}_1}$ leading to an increase in the $Y_{x^{\mathcal{M}}_1}$ value. This is a direct consequence of the multilayering of the layers indicating that by rewiring (`dynamics of the networks') one can change the value of the PEV entries, i.e. `dynamics on networks'  \cite{neuroscience_2017}. In other words, our framework is useful in connecting `dynamics on' and `dynamics of' networks for MNs in terms of the PEV localization. 

Next, we attempt to understand the sensitivity of the PEV in the critical region ($r_3$) for the both-layers rewiring protocol and in the absence of the same in the single-layer protocol. We can witness that for the case of single-layer rewiring, during the evolution of $\lambda_1$ and $\lambda_2$, both show an increase and remain separated to each other (Fig.~\ref{fig6}(c)). However, for the both-layers rewiring protocol, as evolution progresses, $\lambda_2$ starts shifting towards $\lambda_1$, (Fig.~\ref{fig6}(d)) as a consequence of $\lambda_2$ drifting away from the bulk region \cite{SJ_Sanjiv_pre_gev_2014}. This drift in $\lambda_2$ is not surprising as we know that the final optimized structure obtained from the both-layers rewiring consists of two parts in both-layers of $\mathcal{M}_{opt}$. We can observe from Fig. \ref{single-both}(a) that there exist two communities (surrounded by a dotted ellipse) such that for each community one part resides in $\mathcal{L}_1$ layer and another part of the community belongs to $\mathcal{L}_2$ layer of the MN. Hence, there should be two eigenvalues which lie outside the bulk. However, the interesting observation is that for the $\mathcal{M}_{opt}$ obtained from the both-layers rewiring, $\lambda_2$ not only drifts away from the bulk but becomes very close to $\lambda_1$, in fact, $\lambda_1 \sim \lambda_2$. Almost same value for both the eigenvalues might be a reason behind the sensitivity of the PEV  \cite{evec_localization_2017} for the both-layers rewiring as there exist now two eigenvectors in the same eigenspace. Note that, for the single-layer rewiring protocol (Fig. \ref{single-both}(b)) it is hard to get two communities as one layer is fixed which prohibits $\lambda_2$ being separated from the bulk. Hence, there is no possibility of $\lambda_2$ being close to $\lambda_1$ which is always well separated from the bulk of the sparse networks.  

\begin{figure}[t]
\centering
\includegraphics[width=0.95\columnwidth]{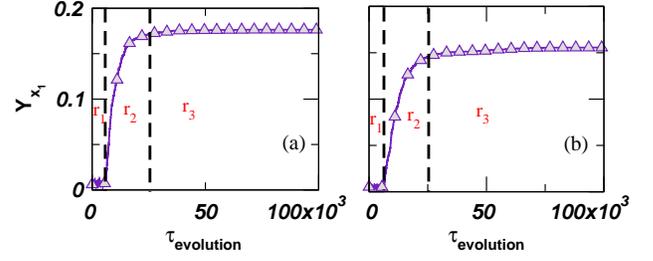}
\caption{(Color online)
The IPR value ($Y_{x_1}^{\mathcal{M}}$) of (a) three layers and (b) four layers MN. The simulation is performed for (a) three and (b) four layers rewiring protocols. Each layer of $\mathcal{M}$ contains $100$ nodes and $\langle k_{\alpha} \rangle=5$.}
\label{fig5}
\end{figure}

Furthermore, we present the results for three layers and four layers MNs (Fig. \ref{fig5}). Starting with the three and four layers initial random MNs, we evolve them using the optimization technique as described above. Again, the optimized rewiring leads to an increase in the IPR value of the MN during the evolution (Fig. \ref{fig5}) with the existence of $r_1$, $r_2$ and $r_3$ regions. For the three layer MNs, we can adopt the rewiring protocol in various manners, (i) rewiring only one layer by fixing other layers, (ii) rewire two layers and fix one layer, (iii) rewire all the layers independently. All the three ways of the rewiring yield the network properties similar to those obtained for the two layers MN in the optimized state.
\begin{table*}[t]
\centering
\begin{tabular}{|l|c|c|c|c|c|c|c|c|c|c|c|c|c|c|}
\hline
\multirow{2}{*}{}\cellcolor{red!25}Network&\cellcolor{green!25}$l$ & \cellcolor{green!25}$N$ & \cellcolor{green!25}$\langle k \rangle$ &\cellcolor{green!25} $Y_{x_{1}^{\mathcal{M}}}$ &\cellcolor{green!25}$k_{max}$ &\cellcolor{green!25}$\langle CC \rangle$&\cellcolor{green!25}$r_{deg-deg}$ &\cellcolor{green!25} $\lambda_{1}$ &\cellcolor{green!25}$\lambda_{2}$&\cellcolor{green!25}$r_{pev-deg}$&\cellcolor{green!25}$r_{pev-cc}$ &\cellcolor{green!25} $\gamma$&\cellcolor{green!25} $\lambda_{1}^e$\\  
\hline
Moscow Athl.  &3 &124423&4.01&0.03&4840&0.08&-0.13&75.22  &71.5&0.66&0.1&2.11&72.26\\
NYClimate& 3 &148936&5.39&0.07  &9742 &0.08 &-0.1 &118.5  &99.2&0.65&0.1&2.19 &101.16\\ 
MLKing2013 &3&318962&2.51 &0.08  &8689 &0.02  &-0.11 &93.2&85.5&0.25&0.0&2.02 &95.92\\ 
Cannes2013 &3&573353&3.98 &0.2  &8676  &0.07 &-0.1 &94.26&86.9&0.38&0.0&2.17 &94.59\\ 
Higgs mux&2&886744&32.57 &0.003 &76034  &0.09  &-0.1& 653.5&436.7&0.71&0.15&2.33 &279.94\\ 
ObamaIsrael&3&2258678&3.55 &0.15  &21650&0.07  &0.0&151.77&139.9&0.43&0.0&2.25 &148.44\\ 
CKM &3&329 &5.95 &0.02 &25 &0.15 &0.1 &7.84 &5.75 & 0.78& -0.16 &3.5 &5.02\\ 
\hline
\em{Drosophila} &4 &10255 &7.62 &0.008 &175 &0.09&0.1 &46.96 &31.0&0.6&0.28&3.5 &13.98\\ 
\em{Homo} &4 &34363 &10.22&0.09 &9570 &0.16&-0.05&118.76 &67.2&0.7&0.11&2.69 &99.70\\
\em{Arabidopsis} &4 &8163 &4.45 &0.24  &1296 &0.07 &-0.1 &36.28  &23.03 &0.6 &0.0 &2.65 &36.90\\
\em{HumanHIV1} &2 &1138 &2.48& 0.24&250 &0.02&-0.45&15.87 &14.93&0.5&-0.05&2.78 &17.97\\
\em{Celegans-nrl.} &3 &791 &9.74 &0.025  &82 &0.18 & 0.11& 21.18 & 13.53 & 0.8 &0.1 &3.18 &9.39\\
\em{Mus} &4 &9657 &4.22 &0.03  &368 &0.06 &-0.16 &34.56  &24.57 &  0.46 & 0.0 &2.59 &20.04\\
\em{Plasmodium} &3 &1161 &4.15 &0.03  &83 &0.02 &0.0 &13.12  &8.75 &  0.8&0.17&3.5 &9.52\\
\em{Rattus} &4 &2906 &2.98 &0.23  &814 &0.07 &-0.14 &29.16  &14.14 & 0.74& 0.12  &2.75 & 30.22\\
\em{SacchCere} &4 &20482 &17.37 &0.02&3187  &0.22 &-0.1 &110.81 &70.51 & 0.65 & 0.11 &2.65 &57.55\\
\em{SacchPomb} &4 &6401 &8.62 & 0.06 &1021 &0.16 &-0.14 &47.95  &36.62  &0.57 &0.12 &2.44 &33.10\\
\hline
\end{tabular}
\caption{Various properties of real-world social (first 7) and biological (last 10) multilayer 
networks. Inverse participation ratio ($Y_{x_1}$), maximum degree ($k_{max}$), average clustering coefficient ($\langle CC \rangle$), degree-degree correlation ($r_{deg-deg}$), PEV-degree correlation ($r_{pev-deg}$), PEV-cc correlation ($r_{pev-cc}$), the largest eigenvalue ($\lambda_1$), 
the second largest eigenvalue ($\lambda_2$), power-law \cite{power_law_2009} scaling parameter ($\gamma$), and  $\lambda_1^e$ \cite{Goltsev_prl2012} of few real-world MNs. Ref.~\cite{implement_python}
is used to calculate IPR and eigenvalues of MNs having large network size. The IPR values of the corresponding random networks are very close to $3/N$ which is predicted by the random matrix theory  \cite{random_matrix}. First six networks with $l$ layers are constructed based on the Twitter data with different exceptional events ranging from sports, politics to scientific discovery of Higgs boson. The layers represent retweet, reply, and mention on twitter \cite{social_networks} network. The CKM is a multilayer social network of a sample of physicians in US \cite{CKM}. The Drosophila and Homo are the multilayer genetic and protein interaction networks where layers are the physical association, direct interaction, colocalization, association respectively \cite{genome_data_1,genome_data_2, genome_data_3}. The rest of the networks are also multilayer genetic and protein interaction networks where we consider only the first four layers when the number of layers is more. We consider the largest connected component to calculate various properties 
and treat all the edges undirected and unweighted. 
}
\label{table2}
\end{table*}

\subsection{Localization in real-world multilayer networks}
Furthermore, we examine the PEV localization of many real-world MNs. We find that the real-world MNs have the PEV which is much more localized than the corresponding random MNs, however, much less localized than the optimized multilayer structure. We present results for MN of the Twitter data collected during the occurrence of different exceptional events like the discovery of Higgs boson in 2012, Cannes Film Festival, the 14th IAAF World Championships in Athletics held in Moscow 2013, Martin Luther King's famous public speech celebrating 50 years ``I have a dream'' in 2013, official visit of US President Barak Obama in Israel in 2013 \cite{social_networks}, a large-scale event on global climate change in New York in 2014. The choice of the Twitter network data provides a good proxy for the large population of the social behaviors \cite{social_networks}. The individual layers of the Twitter MN follow the power-law degree distribution and reflect scale-free topology. In addition to these social networks, we consider biological MNs as well. The multilayer gene-interaction networks, Drosophila and Homo-genetic \cite{genome_data_1,genome_data_2, genome_data_3} consist of layers denoting the physical association, direct interaction, colocalization, and association respectively. We make crude approximations that all the networks are undirected and unweighted. Table \ref{table2} presents PEV localization and structural properties of these MNs. All the networks have IPR value much larger than the corresponding random networks. 

We can estimate the IPR value of a random MN consisting of layers of size $N/l$ represented by ER random network as $Y_{x_{1}^{\mathcal{M}}} \approx 3/N$ \cite{random_matrix}. Other structural properties of such random MN can be calculated as $\langle CC \rangle \approx \frac{\langle k \rangle}{N}$, and  $r_{deg-deg} \approx 0$ for large $N$ \cite{network_sec_2016}. The real-world multilayer networks considered here comprise of structural properties which differ considerably from the corresponding random MNs. From Table \ref{table2}, it is clear that all the real-world multilayer networks considered here contain a hub node having a very large degree. Additionally, they have higher average clustering coefficient value ($\langle CC \rangle$) and smaller degree-degree correlation than the corresponding random networks. Additionally, PEV of these MNs is more localized than the corresponding random MN. From the Table \ref{table2}, it is evident that the less localized networks possess high $r_{pev-deg}$ value, and for most of the real-world MN the $r_{pev-cc}$ value is positive. 
Although these are not very surprising observations, by combining the comparison of measures of various structural properties and IPR values of the real MNs with those of the model MNs achieved during the optimized evolution process, it is evident that real-world MNs lie well above the $r_1$ region. Furthermore, Table \ref{table2} depicts that the largest and the second largest eigenvalue of the real-world MNs are well separated from each other indicating that these real-world MN lie below the $r_3$ region. Note that in the $r_3$ region, the largest and the second largest eigenvalues lie very close to each other leading to the sensitivity of the IPR value of the PEV to single edge rewiring. Based on these two sets of observations we can fairly conclude that the real-world multilayer networks lie in the $r_2$ region of the evolution process of the model multilayer networks.

Additionally,
we calculate the power-law exponent ($\gamma$) \cite{power_law_2009} of the real-world MNs. 
Goltsev et al. \cite{Goltsev_prl2012} states that PEV localization at a node with degree 
$k_{max}$ occurs if estimated $\lambda_1^e=k_{max}/\sqrt{k_{max}-(\langle k^2 \rangle/ \langle k \rangle-1})$
is close to the largest eigenvalue of the network.
We find that those real-world MNs, considered here, which have highly localized PEV obey this relation between the largest eigenvalue and the degree sequence.
However, we can not conclude more on the relation between
the localization properties of the real-world MNs and $\gamma$ value which requires
further rigorous investigation by considering various network sizes and scaling parameters.

\section{Conclusion} In this work, we explore the impact of the optimized rewiring for the PEV localization in MNs. 
We construct MN structures through an optimization process that yields highly localized PEV quantified by the IPR value. 
Our approach provides a comprehensive way to investigate not only the properties of the optimized multilayer structure but also the intermediate multilayer networks before the most optimized structure is found. In other words, we develop a learning framework to explore the evolution of the eigenvector from a delocalized to a highly localized state. We analyze several structural and spectral properties during the network evolution process for the single-layer as well as all the layers of the MNs. For both the protocols, we find that there is an emergence of various structural features as PEV gets localized. Moreover, for both the protocols, there is a noticeable difference present in the spectral properties in the saturation region. For both-layers rewiring protocol, in the saturation region, PEV is sensitive to a single edge rewiring as also observed for the optimized evolution of the monolayer networks. However, interestingly, we get rid of the sensitivity in the PEV in the saturation region by implementing a single-layer rewiring of the MN. Additionally, we have investigated the PEV localization behavior of several large empirical MNs constructed using the data ranging from social to biological systems. Our analysis reveals that these real-world MNs are much more localized than the corresponding random MNs, and also have structural properties close to those obtained in the $r_2$ region of the optimized evolution process of the model MNs. 
Further, we show that by rewiring a single-layer, one can tune the contribution of the node weights of the other layer to the PEV of the entire MN. 
Rearrangement of the node weights used in semi-supervised based learning and has great practical importance in machine learning \cite{semi_supervised_learning_2017}. 

This work can be extended to confine or facilitate propagation of perturbation in a network by an appropriate multilayering with other networks. For instance, in the case of a disease outbreak in a 
society, which already has a connection network among its people, one can create another 
network (for instance virtual-world awareness network) comprising of the same people in a manner 
such that the PEV of the whole MN gets localized leading to a restriction of the disease in a 
small section of the society. Furthermore, there exist several open questions which
require future investigations. An important direction is to understand the pre-requisite of the second largest eigenvalue becoming very close to the largest one in the most localized structure. Additionally, the current article focuses on the localization of only one eigenvector, and it will be interesting to characterize network properties leading to more than one localization points. Furthermore, this article considers only random edge rewiring adopted during the optimization
process. It will be interesting to see the consequences of restricted rewiring such as 1-k, 2-k rewirings. \cite{systematic_topology_analysis_2006}.

{\bf Acknowledgement:} SJ acknowledges DST, Govt of India grant (EMR/2014/000368 and EMR/2016/001921) for financial support and PP thanks to DST (EMR/2014/000368) for the JRF fellowship. We thank Manlio De Domenico (Universitat Rovira i Virgili) for helping with multilayer network datasets. 
SJ is indebted to  M. V. Ivanchenko (Lobachevsky State University of Nizhny) and Charo I. del Genio
(University of Warwick) for useful discussions, and M. S. Baptista (University of Aberdeen, UK) 
and Manavendra Mahato (IIT Indore) for carefully reading the manuscript and suggesting changes. 
PP thanks members of CSL at IIT Indore for discussions.

\end{document}